# First Principal Analysis of Novel Half Heusler Alloys VPdZ (Z= Ge, Sn) for thermodynamic, spintronics and optoelectronic applications.

Ashwani Kumar[1], Shyam Lal Gupta[2], Sumit Kumar[3,] Anupam[4] and Diwaker[5, *]

[1]Department of Physics, School of Basic Sciences, Abhilashi University Mandi-175045 (INDIA)
[2]Department of Physics, Harish Chandra Research Institute, Allahabad- 211019 (INDIA)
[3]Department of Physics, Govt. College Una-174303 (INDIA)
[4]Department of Physics, RGM Govt. College Jogindernagar-175015 (INDIA)
[5*]Department of Physics, SCVB Govt. College Palampur-176061 (INDIA)

*Email: diwakerphysics@gmail.com

## Abstract

The study explores the structural stability, elastic, mechanical, electronic, thermophysical, magnetic, optical and lattice dynamic properties of VPdZ (Z= Ge, Sn) half Heusler alloys using density functional theory. The alloys show stability in type-α ferromagnetic phase and have half-metallic band topologies. The half-metallic feature is indicated by the spin-polarized behavior that is revealed by the detailed profiles of the electronic band structures. According to the electronic band profiles both alloys are half-metallic, with indirect energy gaps in the spin down channel of 1.10 eV and 1.02 eV for VPdGe and VPdSn half Heusler alloys respectively. The Quasi-Harmonic Debye model helps study thermodynamic parameters, and the magnetic moment values align with the Slater-Pauling rule. The alloys have potential applications in thermodynamic, spintronic, and optoelectronic fields.

## 1. Introduction

Past few decades the world is confronting major energy crisis which is profoundly affecting the various countries. With significant upsurge in energy consumption, steady reduction and enormous carbon emission of the limited non-renewable energy resources is one of primary cause. Several modern technologies pertaining to energy transformation and storage from the conventional sources have become the centre of the study for many researchers. The researches around the globe inevitably focussing to develop the multifunctional materials based on chalcogenides [1,2], organic compounds [3,4], perovskites [5,6], topological insulators [7,8] and half Heusler alloys [9,10]. Thermoelectrics and spintronics are the twigs of half-metallic ferromagnetism (HMF). Mostly HMF devices based on the inconsistent number of minority and majority spin carriers and exhibited 100% spin polarization near the Fermi level ($E_F$). The half metallic materials generate a fully spin-polarized current, which enhance the efficiency of magneto-electronic devices. Such materials show the hybrid properties of metals and semiconductors and are known as half-metallics. Half-metallic (HM) materials exhibit spin-dependent electronic band polarisation in which metallic in one orientation whereas semiconducting in another one. Heusler alloys are the incredible materials with HM and other multifunctional properties utilised in spintronics, shape memory effects, spin gapless semiconductors, thermoelectricity and optoelectronics [11-15]. The reason to delegate Heusler alloys in the mentioned field is due to their unique crystal structure as well as electronic band profile and other tunable properties. Transition metal based Heusler alloys (HAs) have been extensively investigated by material researchers worldwide. Predominantly, the vanadium (V) based Heuslers establish a huge family with half metallic or semiconducting band profiles. In the field of spintronics, these magnetic materials have found significant applications.

Over the years many researchers have studied Heusler compounds theoretically [16-18] and experimentally. Through the experimental investigations, Groot et al. [19] found the first HM characteristics in a NiMnSb half-Heusler. Later on, it was projected that $Fe_2TiAs$ and $Fe_2TiSb$ are the half-metallic (HM) full-Heuslers [20]. Furthermore, HM behavior in $Co_2RhSi$, and $Co_2RuSi$ full-Heuslers was theoretically predicted by A. Kumar et al. [21]. The HM behavior was computationally proved by Djelti and co-workers [22]. Dubey et al. [23] noticed HM ferromagnetic activity when examining the elastic, electrical, and structural behavior of FeCrAs. Additionally, Zahir et al. [24] reported on their first-principles calculations of the CoFeZ (Z = P, As, and Sb) half-Heuslers. To fulfil the demands of contemporary technological applications in spintronics, thermoelectrics, electronics, and other new and unexplored domains, it is imperative to find unique HHs. Further investigation into the physical characteristics of half-Heuslers is necessary to determine their viability and broad applicability. In this work, we report on the investigation of VPdGe and VPdSn half Heusler alloys by the application of Boltzmann transport theory and quantum density functional theory (DFT). In order to guarantee the mechanical stability of the optimized crystal unit cells, the crystal structures of the suggested Heusler alloys were adjusted. The optimized structures' phonon dispersion relations were calculated, confirming their thermodynamic stability. Even at temperatures as high as 1000K, the thermodynamic stability of both compounds is determined by ab-initio simulations. Numerous physical parameters, including structural, dynamical, electronic, optical, thermodynamic, and lattice dynamic properties, were computed in order to look into the potential applications of these two full Heusler alloys.

## 2. Computational detail

In the current study, we used density functional theory to investigate the dynamical stability, electronic properties, optical, thermodynamic and magnetic response and lattice dynamic stability of VPdGe and VPdSn Heusler alloys. Density functional theory well-known for its accuracy and dependability, guarantees the high correctness of different characteristics. The Kohn-Sham equations are successfully solved using the full potential linearly augmented plane wave (FP-LAPW) basis set in the WIEN2k simulation program [25]. Within the bounds of the generalized gradient approximation (GGA), which is precomputed by the Perdew-Burke-Ernzerhof (PBE) exchange-correlation potential, the electron-electron interaction is handled superbly [26]. In the calculations, GGA is successfully executed to obtain the detailed electronic structure, structural stability and magnetic response. The structural unit cells of VPdGe and VPdSn alloys are divided into two portions to initiate the simulations: the interstitial space and the muffin-tin sphere. One popular approximation technique that is frequently used to determine the energy state of an electron inside a crystal lattice is the muffin-tin approach. Muffin-tin radii (RMT) for V, Pd, Ge and Sn atoms are 2.18, 2.23, 2.24 and 2.25 respectively. Outside the sphere, the electrons' wave functions are expressed as plane waves with a wave vector cut-off of $R_{MT} \times K_{max}=8.0$, where $K_{max}$ denotes the reciprocal lattice vector. A constant energy of magnitude -7.0 Ry is used to distinguish the core and valence electrons, while a fixed energy value of $10^{-5}$ Ry is effectively employed to achieve the scf convergence. The Brillouin zone for the present system is sampled using the Monkhorst pack [27] technique with a (14 × 14 × 14) k-point mesh. IRelast [28] in the WIEN2K package is used to investigate elastic parameters. The lattice dynamic stability and thermodynamic response are investigated using the Phonopy [29] and BoltzTrap2 codes [30].

## 3. Results and discussions

### 3.1 Structural properties

The regular Heusler relaxed in centrosymmetric cubic *fcc* with L$_{21}$ structure, *fm-3m* space group no. #225 respectively, whereas the half or semi Heusler alloys (HAs) mostly crystallize in non-centrosymmetric cubic *fcc* with structure type C$_{1b}$ and *f-43m* with space group no. #216. The rock salt (RS) sublattice of YZ serves as the foundation for the crystal structure of the α-phase half Heusler VPdZ (Z; Ge, Sn) system. AY or YZ form the sublattices in the XYZ type structure. The magnetic ordered crystal structure for VPdZ (Z; Ge, Sn) is given in schematic **Fig. 1 (a) & (b).**

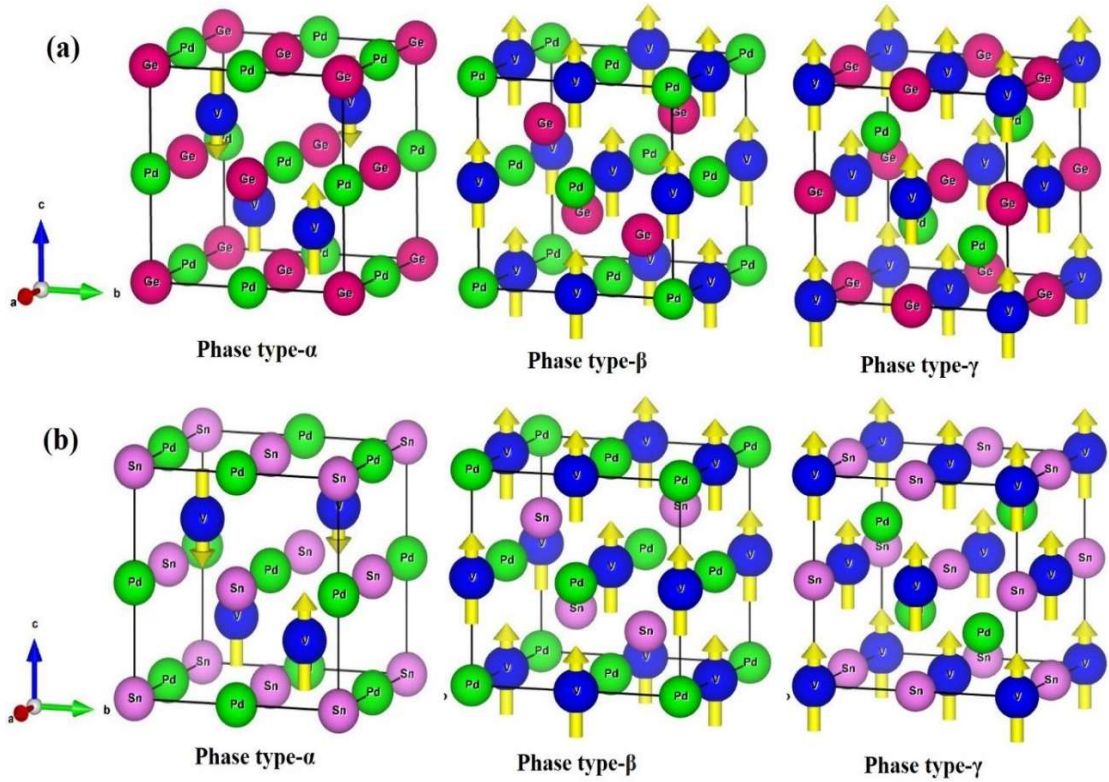

**Fig.1:** Crystal structures in *F-43m* symmetry for (a) VPdGe and (b) VPdSn vanadium metal based half Heusler alloys in phase type- α, β and γ.

The **Table 1** present the appropriate Wyckoff positions and three potential phase type-α, β, and γ.

**Table 1:** Wyckoff's positions for VPdZ (Z; Ge, Sn) half Heusler alloys in α, β and γ-phase types

| Structure | Li/Na | V | Sb |
|---|---|---|---|
| α | (1/4, 1/4, 1/4) | (1/2, 1/2, 1/2) | (0, 0, 0) |
| β | (1/2, 1/2, 1/2) | (0, 0, 0) | (1/4, 1/4, 1/4) |
| γ | (0, 0, 0) | (1/4, 1/4, 1/4) | (1/2, 1/2, 1/2) |

The considered alloys VPdZ (Z; Ge, Sn) alloys display a structure $C1_b$ exhibit three dissimilar atomic arrangements identified as to α, β and γ phases. It is found that the atomic locations in both crystal structures determine the dynamical stability. So, it is inevitable to find the most stable phase with the least amount of energy among the current phases. The computation process shows that, in comparison to type-β and type-γ, the type-α structure of both VPdGe and VPdSn alloys is exhibit more stability as given in **Table 2**. In the ferromagnetic (FM) phase, the both crystal structures displays robust stability than that of the non-magnetic (NM) phase. We calculated the mean lattice parameters and lowest energy values for VPdGe and

VPdSn alloys. **Eqn. 1** expresses the successful execution of the Murnaghan equation of the states [31] for volume optimization in the ferromagnetic state.

$$E_T(V) = E_0 + \frac{B_0 V}{B_0'(B_0'-1)}\left\{\left(\frac{V_0}{V}\right)^{B_0'} - 1) + B_0'\left(1 - \frac{V_0}{V}\right)\right\} \quad (1)$$

where $V_0$ represent the unit cell's volume at zero pressure, $B_0$ is the bulk modulus and $E_0$ signifies the minimal total energy. The computed lattice constant ($a_0$) with GGA scheme for VPdGe and VPdSn at symmetry alongwith the other parameters $V_0$, $B_0$, and $B_0'$ are provided for the first time in **Table 2**.

**Table 2:** Calculated crystal structure stability parameters.

| Alloy | Type | $a_0$(Å) | | $B_0$(GPa) | | Volume (a.u.$^3$) | | Energy (Ry) | |
|---|---|---|---|---|---|---|---|---|---|
| | | FM | NM | FM | NM | FM | NM | FM | NM |
| VPdGe | α | 5.99 | 5.87 | 137.82 | 143.80 | 345.02 | 342.27 | -16190.937822 | -16190.926515 |
| | β | 5.98 | 5.86 | 138.56 | 144.12 | 346.21 | 343.23 | -16190.926754 | -16190.913615 |
| | γ | 5.97 | 5.88 | 137.87 | 143.98 | 347.32 | 344.15 | -16190.916532 | -16190.910653 |
| VPdSn | α | 6.14 | 6.13 | 118.57 | 125.48 | 391.65 | 389.11 | -24351.003877 | -24350.991726 |
| | β | 6.23 | 6.14 | 117.65 | 124.43 | 392.43 | 390.25 | -24351.003121 | -24350.954352 |
| | γ | 6.43 | 6.13 | 116.45 | 123.87 | 393.42 | 389.92 | -24351.003006 | -24350.965143 |

**Fig. 2(a)** and **(b)** shows the volume versus energy plot. For both alloys, the ferromagnetic (FM) state is relatively stable than the non-magnetic (NM) state.

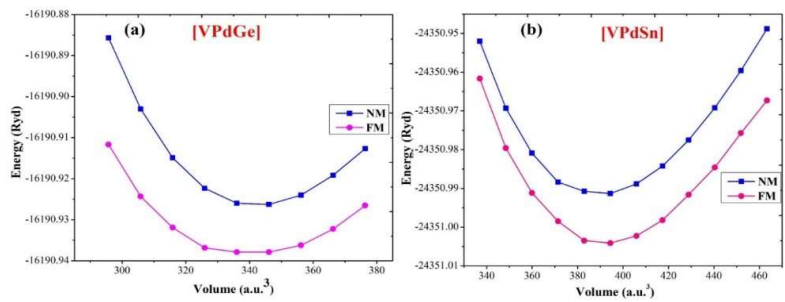

**Fig.2:** Volume optimization curves, **(a)** Ferromagnetic (FM) and **(b)** non-magnetic (NM) with stable phase type-α for VPdZ (Z; Ge, Sn) Heusler alloys.

### 3.2 Electronic properties

The electronic band structure and density of states (DOS) are vital for assessing the electronic behavior of materials. The forbidden energy gap ($E_g$) is sited between the empty conduction band (CB) and full valence band (VB) in the band structure. The band structure plots energy gab region regulates the material's thermoelectric response. We have precisely determined the band gap of VPdZ (Z; Ge, Sn) alloys using the GGA-PBE method. **Fig. 3(a)** displays the band structure for both spin-up and spin down mode. We may observe that for spin up orientation there is overlapping of conduction and valence band at different symmetric point which clearly signifies that majority of spin bands in this orientation are of metallic character.

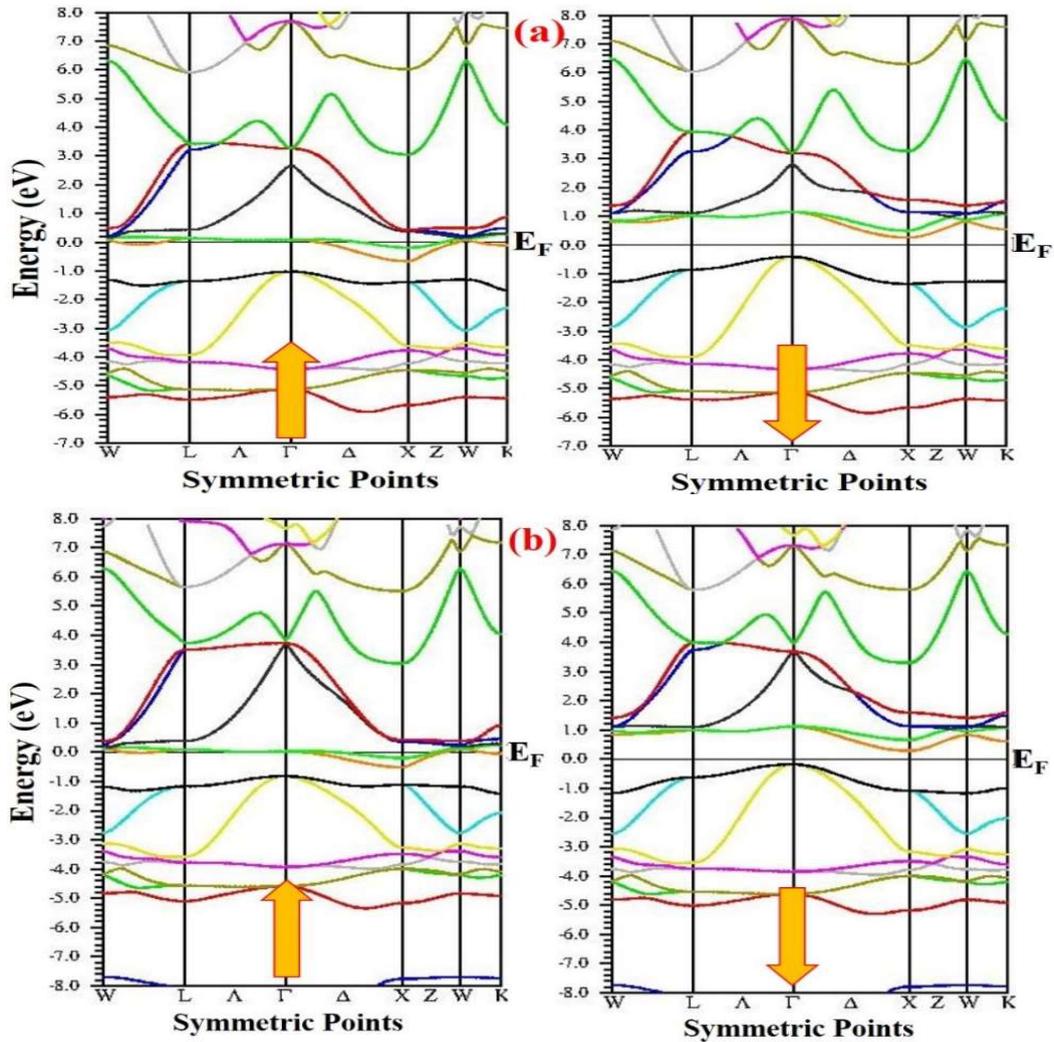

**Fig. 3:** Electronic band structure of VPdZ (Z; Ge, Sn) for **a)** spin up and **b)** spin down configurations

In case of spin-down channel we observe an indirect band gap of 0.72eV and 0.62eV alloys at X symmetry point in band structure diagram for both alloys consequently, exhibit the semiconducting behaviour as shown in **Fig. 3(b)**. The deepest point of the conduction band lies at the X equilibrium point, whereas the highest point of the valence band lies at the gamma (Γ) symmetric points. This is due to the exchange interaction among the electrons of V, Pd and Ge,

Sn sp main block metal atoms which causes spin up and down states to split, where spin down states needs less energy results semiconducting or insulating behaviour. Another reason is that the spin-down channel has antibonding states that sit above the Fermi level while the bonding states stay filled, creating a semiconducting gap for spin down channel depending on the relative energy levels and occupation.

The detailed total density of state (TDOS) and projected density of states (pDOS) plots of VPdZ (Z; Ge, Sn) alloys in both spiral (up and down) directions are provided in **Fig. 4 (a)** to **(f)**. These graphs provide a thorough understanding about the electrical behavior. The precise contribution of V, Pd, Ge, and Sn atoms electrons in given material is explained by carefully examining pDOS.

The exchange splitting that occurs between the V & Pd, Pd and Ge & Sn metal atoms is primarily liable for the half-metallic nature of both materials. The *d*-orbital states of V and the p-orbital states of Pd atoms are held responsible for most of the DOS observed at the Fermi energy, according to pDOS. In contrast, the s and p orbital states of Ge and Sn atoms contribute to the electronic behavior, whereas the p-states of Pd atoms moderately contribute to both spin channels. The electrical behavior for both spin orientations is largely influenced by the d-states of V and Pd atoms. The d-states of the V atom are significantly placed in the conduction band in the minority spin state.

$$P = \frac{DOS_\uparrow(E_F) - DOS_\downarrow(E_F)}{DOS_\uparrow(E_F) + DOS_\downarrow(E_F)} \times 100 \qquad (2)$$

$DOS_\uparrow(E_F)$ and $DOS_\downarrow(E_F)$ in Eqn.10 denotes the energy states for minority and majority spin channels in the $E_F$.

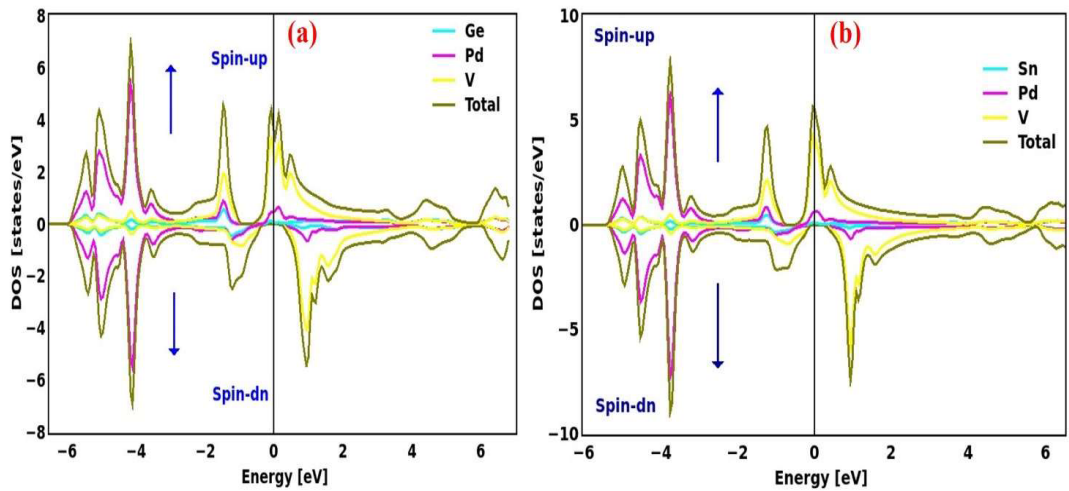

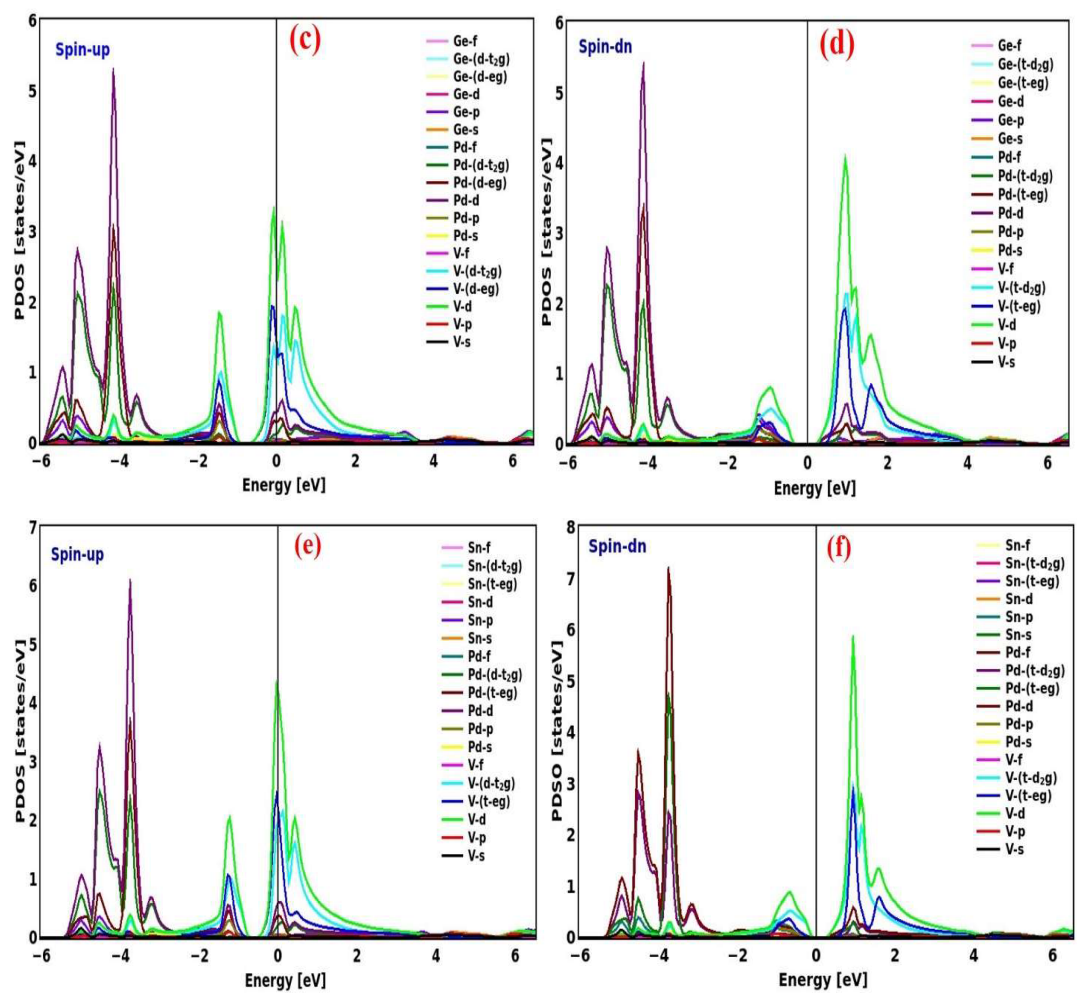

**Fig. 4:** Total and projected density of states **a) & b)** TDOS, **c) & d)**, **e) & f),** pDOS for VPdGe & VPdSn alloys in spin up and down configurations

### 3.3 Mechanical stability

By computing the distinguished elastic parameters and establishing the relationship between stress and strain we may know the elastic and mechanical response of given VPdGe and VPdSn alloys. The deformation of the alloys under an external force is determined by the elastic parameters, which then allow the material to return to its initial configuration after removal of this force. By knowing the elastic constants at the optimal conditions is vital for examining the mechanical stability of alloys. The elastic constants are investigated to know the utilisation of alloys in mechanical applications. Since both alloys are stable in phase α cubic symmetry so, from the computed date we are able to determine the values of the three separate elastic parameters, $C_{11}$, $C_{12}$, and $C_{44}$. where $C_{12}$ measures the transverse deformation and $C_{11}$ assesses the longitudinal expansion. To describes the mechanical stability the equilibrium elastic coefficients essentially fulfil the Born's stability criteria. The criteria for cubic symmetry are given as-

$$C_{11} - C_{12} > 0 \text{ and}$$
$$C_{11} + 2C_{12} > 0$$
$$\text{Here, } C_{11} > 0 \ ; \ C_{44} > 0 \tag{3}$$

**Table 3** demonstrates that VPdGe and VPdSn alloys satisfies the stability requirements and indicates the mechanical stability. Furthermore, the stiffness and compressibility are determined by the other stated factors, which are referred to as bulk modulus (B) and shear modulus (G), respectively. The parameters G and B can be used to calculate the restoring force which comes into play against the fracture, or resistance to plastic deformation. The following equations can be used to get these values-

The bulk modulus, B is given as- $\quad B = \frac{C_{11}+2C_{12}}{3} \tag{4}$

Similarly, the shear modulus is determined as- $\quad G = \frac{G_R+G_V}{2} \tag{5}$

The R and V in the subscript of **Eqn. 4** denotes the Reuss and Voigt bounds and their values are determined by using the subsequent relations-

$$G_R = \frac{5C_{11}(C_{11}-C_{12})}{5C_{11}+3(C_{11}-C_{12})} \tag{6}$$

$$\text{and} \quad G_V = \frac{C_{11}-C_{12}+3C_{44}}{5} \tag{7}$$

The cubic crystal symmetry is consistent with the mathematical relationship between the different elastic constants. The Kleinmann parameter ($\xi$) is used to calculate the internal strains related to bond twisting and elongation. It evaluates the bending of a simple connection under outside forces. The calculated value of $\xi$ indicates that the VPdGe and VPdSn alloys exhibit greater resistance to a broader range of stresses, hence validating their use in industrial settings. By using the average sound velocity to calculate the Debye temperature ($\theta_D$), we were able to verify the thermodynamic stability, specific heat at low temperature, and phonon stability.

$$\theta_D = \frac{h}{k}\left[\frac{3n}{4\pi}\left(\frac{\rho N_A}{M}\right)\right]^{1/3} V_m \qquad (8)$$

Here, the longitudinal ($V_l$) and transverse ($V_t$) velocities are used to obtain the average sound velocity, which is expressed in the form given as follows-

$$V_m = \frac{1}{3}\left(\frac{2}{V_s^3} + \frac{1}{V_l^3}\right)^{-1/3} \qquad (9)$$

For VPdGe and VPdSn alloys, the predicted Debye temperature ($\theta_D$) at 5GPa pressure is 479.077 K, while for other alloys, it is 494.740 K. We employed the elastic constant and Fine's equation [32] to get the melting temperature-

$$T_m(K) = [553(K) + (5.911)C_{11}]GPa \pm 300K \qquad (10)$$

The computed melting temperatures for VPdGe and VPdSn at 5GPa are 1742K and 726K, respectively. Elevated melting temperature results validate that both alloys can sustain their ground state structure across a broad temperature range, hence contributing to their stability. One important metric for characterizing a material's brittleness or ductility is the Pugh ratio (B/G). In general, brittle materials have a B/G ratio smaller than 1.76. The calculated B/G ratio values for VPdGe and VPdSn alloys show that the former is ductile and the latter is brittle. **Table 3** present the computed values of the above elastic parameters.

**Table 3:** Computed values of elastic parameters ($C_{ij}$ in GPa), bulk and Young's modulus ($B_v, B_R, B_H, G_S, G_v, G_H, E_Y, E_R, E_H$ in GPa), Poisson's ratio ($\nu$ no units), Cauchy pressure (CP in GPa), transverse, longitudinal and average wave velocity ($v_t$, $v_l$ and $v_a$ in m/s) and Debye temperature ($\theta_D$ in K), Pugh's ratio (k no units), Kleinman's parameter ($\xi$ in GPa), Shear modulus (G in GPa), Lame's 1$^{st}$ & 2$^{nd}$ parameter ($\lambda$ & $\mu$ in GPa) Chen-Vickers hardness ($H^{CV}$ in GPa), Tian-Vickers hardness ($H^{TV}$ in GPa), under external pressure of 5GPa & 10 GPa for cubic VPdGe and VPdSn Heusler alloy.

| Stress Parameters | VPdGe | | VPdSn | |
|---|---|---|---|---|
| | 5 GPa | 10 GPa | 5 GPa | 10 GPa |
| $C_{11}$ | 201.181 | 159.219 | 29.250 | 226.124 |
| $C_{12}$ | 219.263 | 140.274 | -59.962 | -122.952 |
| $C_{11} - C_{12}$ | -18.049 | 18.945 | 89.212 | 103.172 |
| $C_{11} + 2C_{12}$ | 639.526 | 439.767 | -90.674 | 472.028 |
| $C_{44}$ | 391.999 | 227.996 | -98.554 | 785.120 |
| $CP = C_{12} - C_{44}$ | -172.736 | -87.722 | 38.592 | 539.216 |
| $B_v$ | 213.236 | 146.589 | -30.225 | 157.343 |
| $B_R$ | 213.236 | 146.589 | -30.225 | 157.343 |
| $B_H$ | 213.236 | 146.589 | -30.225 | 157.343 |
| $G_S$ | 231.583 | 140.587 | -41.290 | 491.706 |
| $G_v$ | -23.412 | 22.292 | 347.298 | 117.395 |
| $G_H$ | 104.085 | 81.439 | 153.004 | 304.551 |
| $E_Y$ | 510.089 | 319.591 | -85.113 | 722.499 |
| $E_R$ | -72.906 | 63.650 | -368.137 | 282.040 |
| $E_H$ | 268.559 | 206.143 | -667.743 | 555.345 |
| $\nu_v$ | 0.101 | 0.137 | 0.031 | -0.265 |
| $\nu_R$ | 0.557 | 0.428 | -1.530 | 0.201 |
| $\nu_H$ | 0.290 | 0.266 | -3.182 | -0.088 |
| $\xi$ | 2.016 | 1.536 | -1.387 | 0.905 |
| $v_t$ | 3670.650 (m/s) | 3199.659(m/s) | 4317.353 (m/s) | 5994.395 (m/s) |
| $v_l$ | 6750.401 (m/s) | 5663.768 (m/s) | 4601.155 (m/s) | 8153.197 (m/s) |
| $v_a$ | 4094.929 (m/s) | 3558.815 (m/s) | 4404.132 (m/s) | 6459.581 (m/s) |
| $\theta_D$ | 479.077 (K) | 420.441 (K) | 494.740 (K) | 733.423 (K) |
| k | 2.049 | 1.800 | 0.198 | 0.517 |
| $\lambda$ | 143.845 | 92.296 | -132.227 | -45.691 |
| $\mu$ | 104.085 | 81.439 | 153.004 | 304.551 |
| $H^{CV}$ | 10.085 | 10.189 | 250.023 | 119.897 |
| $H^{TV}$ | 10.913 | 10.627 | ---- | 111.770 |

**3.4 Optical properties**

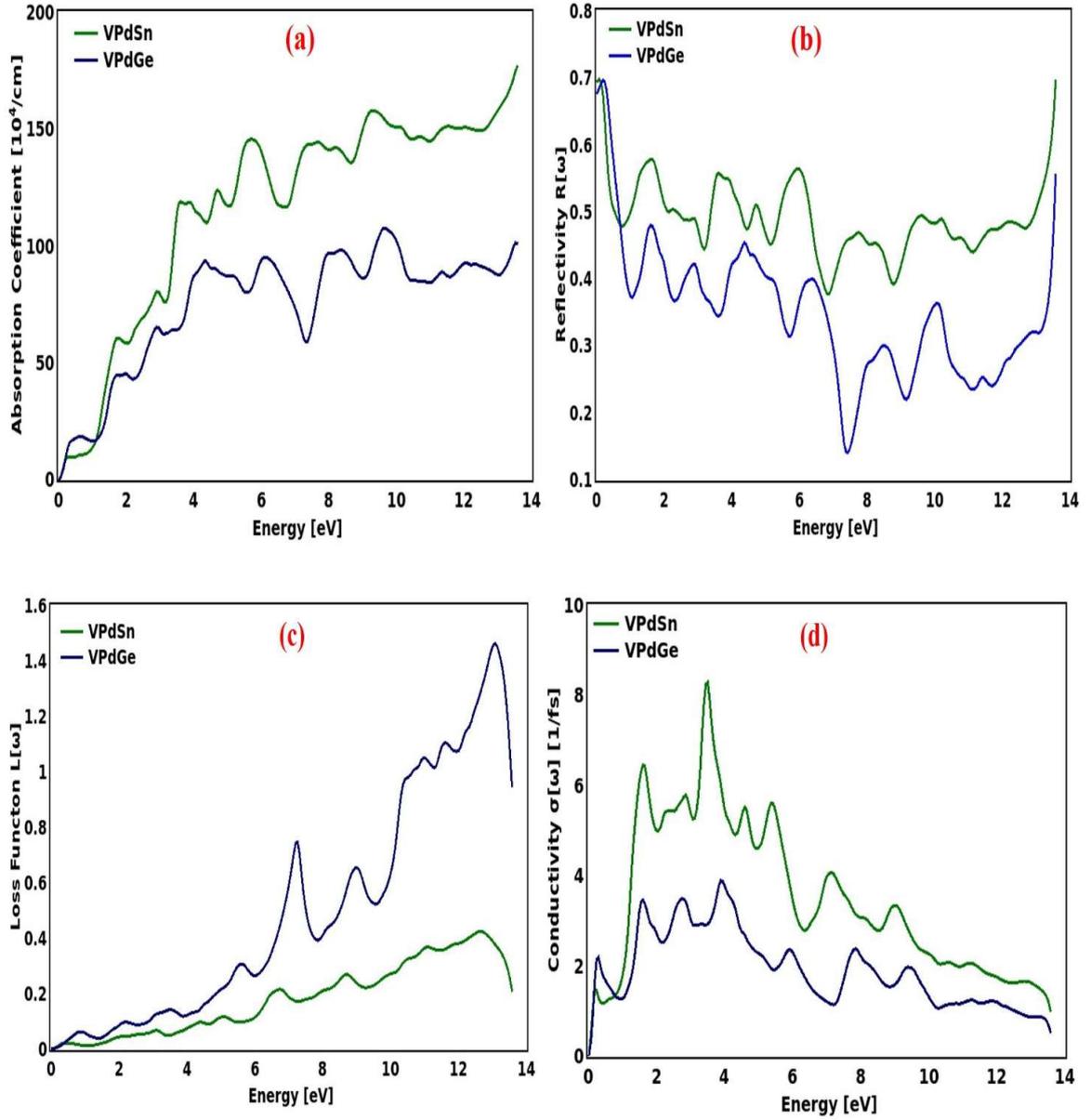

**Fig. 5:** Optical properties for VPdZ (Z=Ge,Sn) HH alloys [a]Absorption coefficient, [b]Reflectivity, [c] Loss function, [d] Conductivity

The focus of this work is on intra-band transitions that disclose metallic behaviour, which greatly boosts optical conductivity in the electromagnetic spectrum's lower energy region. A key component that influences how far light may travel through a material depending on the wavelength of incoming light is the absorption coefficient denoted by $\alpha(\omega)$. Elevated absorption coefficients of materials facilitate the rapid absorption of light photons, therefore inducing the stimulation of electrons in their conduction band. Our computed frequency-dependent absorption coefficients for half Heusler alloys VPdZ (Z = Ge, Sn), are shown in Figure 5 (a) when plotted against the incident photon frequency range of 0-14 eV. The absorption of VPdGe increased noticeably with increasing incident radiation frequency, but remarkably, no absorption occurred when there were no photons on the surfaces of any of these compositions. Based on the plotted graphs, the highest absorption coefficient values VPdZ (Z = Ge, Sn), close to the infrared region are $1 \times 10^5$ and $1.7 \times 10^5$, respectively. Reflectivity is a characteristic that quantifies a material's surface's capacity to reflect incoming radiation. We plotted reflectivity versus incident photon frequency, which was recorded in the range of 0 to 14 electron volts, in order to study the frequency-dependent reflectivity $R(\omega)$ VPdZ (Z = Ge, Sn) HH alloys. Graphs for reflectivity are plotted in Figure 5 (b). The corresponding static reflectivity $R(\omega)$ at zero frequency R (0) for VPdZ (Z = Ge, Sn) is 0.68 and 0.69 respectively. Peak reflectivity for these compositions is 0.5 and 0.7 respectively. The energy loss function $L(\omega)$ expresses the amount of energy lost by scattering or dispersion during an electron's transition. The scattering probabilities that arise during inner shell transitions serve as the basis for the correlation. Figure 5 (c) shows the plotting of the optical loss function versus incident photon frequency for these compositions in the range from 0 to 14 eV. From the graphs we find that the peak optical loss values are 1.5 at a photon energy of 13eV for VPdGe and 0.3 at a photon energy of 12.5eV for VPdSn respectively. In recent few years , the photoelectric effect was used to analyse how incident radiation interacts with the surface of VPdZ (Z = Ge, Sn) HH alloys to break bonds and cause conduction. As illustrated in Figure 5 (d), the conductivity of VPdZ (Z = Ge, Sn) HH alloys as a function of frequency is plotted against the incident photon energy, with range from 0 to 14 eV. The maximum conductivity value was recorded for VPdSn, which reached 8.5 (1/fs) at 4 eV photon energy. VPdSn, owing to its excellent photon conductivity at low energy, is a suitable material for optoelectronic applications, according to the optical evaluation of these compositions.

### 3.5 Thermodynamic properties

To forecast the critical behavior of VPdGe and VPdSn Heusler alloys for several thermodynamic applications, we have implemented the Quasi-Harmonic Debye model [33] at high pressure and temperature. A pressure from 0-10GPa and temperature range of 0-600K. The impact on specific heat at constant pressure and volume ($C_p$, $C_V$), thermal expansion coefficient ($\alpha$), unit cell volume (V), Debye temperature ($\theta_D$), and entropy change (S) has been studied. The computed value of $C_p$ and $C_v$ is plotted as a function of T and P for both VPdGe and VPdSn Heusler systems in the schematic **Fig. 6(a)** to **(d)**.

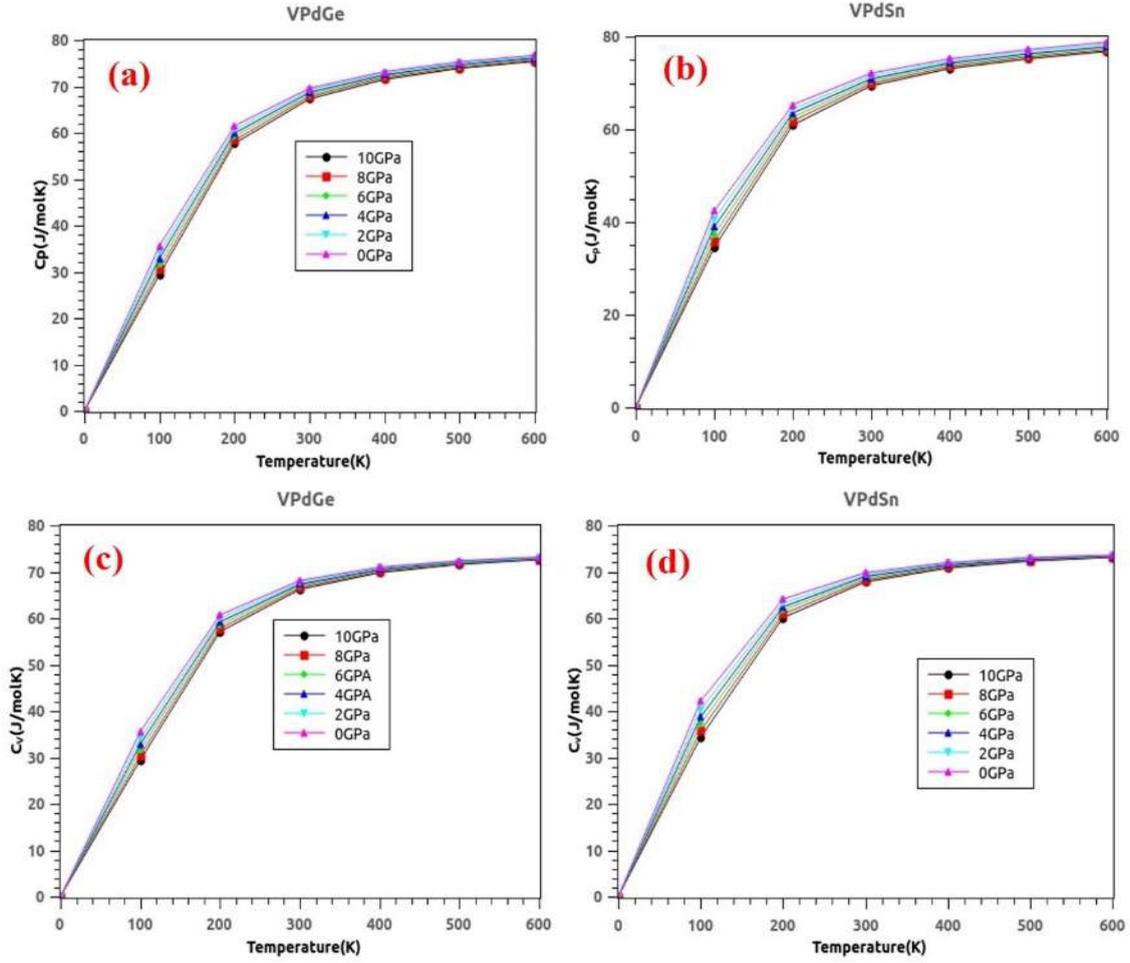

**Fig. 6:** Variation of $C_p$ & $C_v$ with T, **(a)** to **(d)** for VPdGe and VPdSn Heusler alloys

We observed that at low value of temperature, the variations of $C_p$ and $C_V$ are localised for different pressures signifying the dependency on temperature owing to the anharmonic effects. At higher temperatures, the specific heat at constant pressure ($C_p$) and specific heat at constant volume ($C_v$) reaches a constant value and follows the Dulong and Petit's rule [34]. This, as a result of the anharmonic effect which is the same in case of both reported materials at higher temperatures. As a result, we concluded that T affect the both $C_p$ and $C_v$ more than P. Another crucial measure to assess the material's thermal condition is the thermal expansion coefficient (α). It offers important details regarding the solid's anharmonicity and bonding strength. The schematic **Fig. 7(a), (b)** illustrates how α changes with temperature (T) and pressure (P). It is noted that α steadily decreases with significant upsurge with P and T values. The value of α is zero at absolute zero temperature, and it rises rapidly with temperature, suggesting that both alloys obey the Debye's $T^3$ law at lower temperatures.

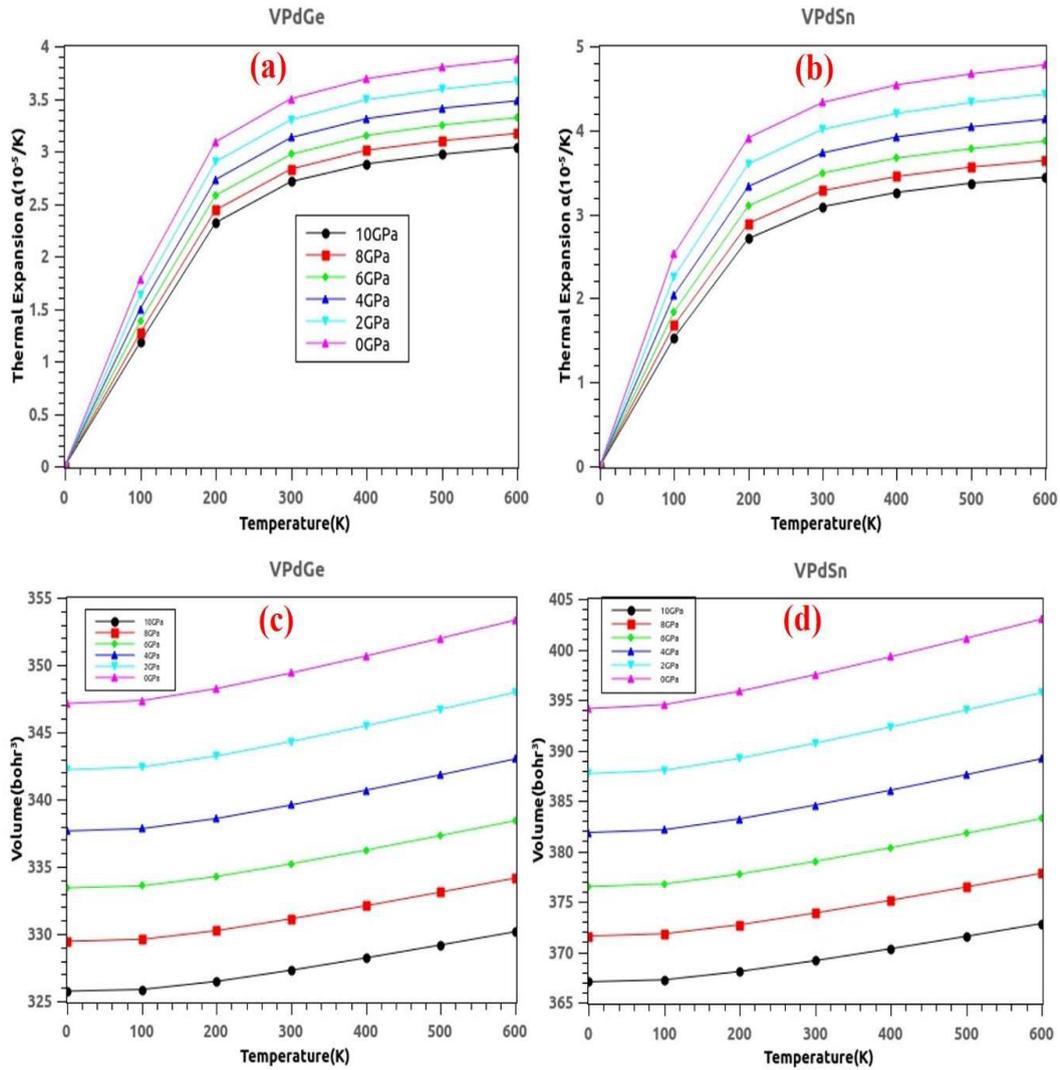

**Fig. 7:** Variation of α & V with T, **(a)** to **(d)** for VPdGe and VPdSn Heusler alloys

Additionally, it is shown in **Fig. 7 (c), (d)** that the volume rises with temperature. On the other hand, the volume rapidly decreases with increase in pressure. Hence, it is confirmed from the observed behaviour that pressure affects volume of both alloys robustly than that of the temperature.

The melting point, heat capacity, and elastic constant are few of the solid-state physical properties that the Quasi Harmonic Approximation (QHA) Debye model associates to $\theta_D$. The predicted dependence of $\theta_D$ on temperature and pressure is displayed in In **Fig. 8 (a), (b)**. The calculated $\theta_D$ at 0K and 0GPa was 420.30K. As the temperature rose and the external pressure remained constant, the $\theta_D$ dropped. Furthermore, we observed that the $\theta_D$ increases with increasing pressure at a constant T.

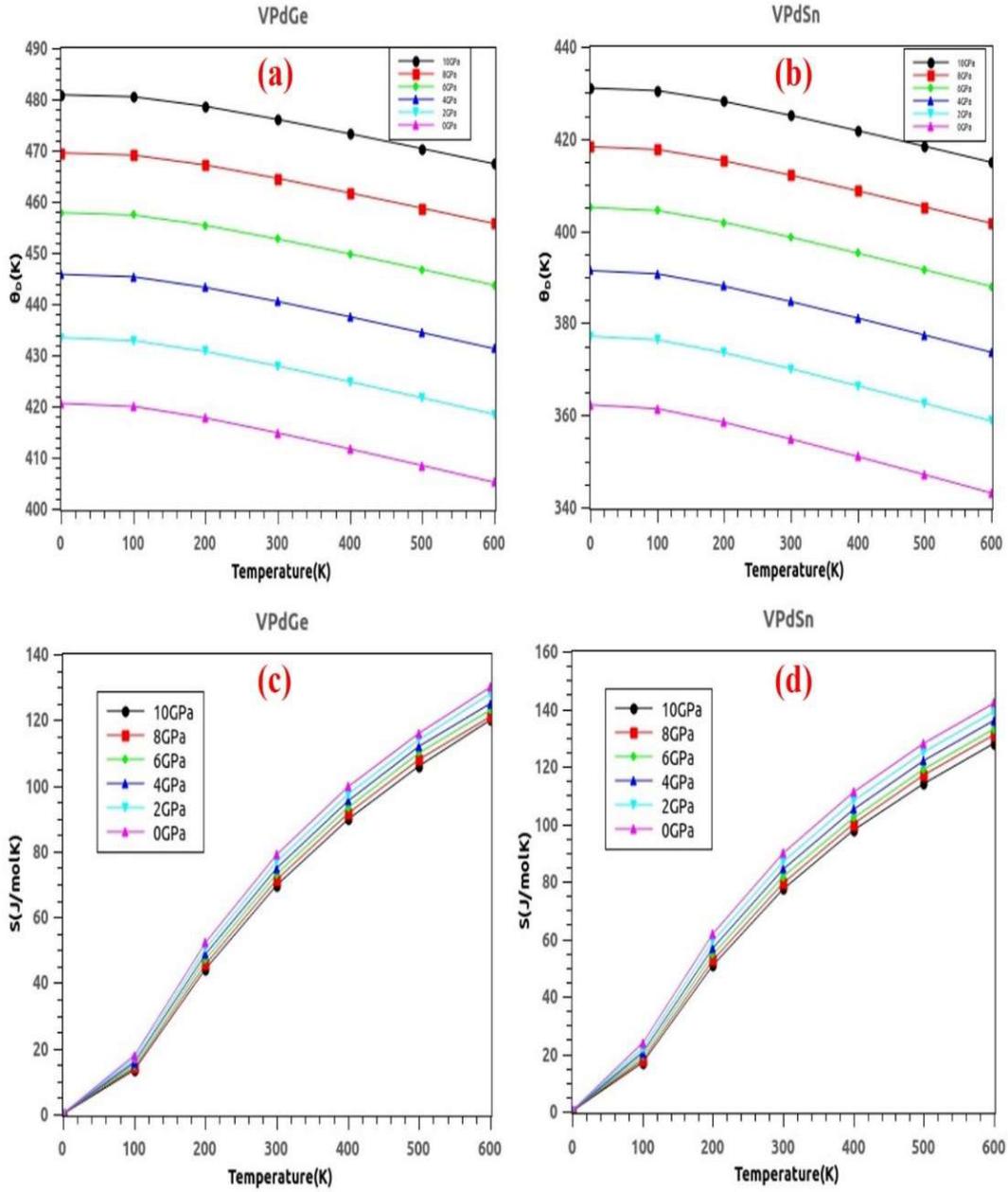

**Fig. 8:** Variation of $\theta_D$ & S with T, **(c)** & **(d)** for VPdGe and VPdSn Heusler alloys

**Fig. 8 (c), (d)** shows the pressure dependence of entropy for VPdGe and VPdSn alloys at various temperatures. The values of entropies rise with temperature and fall with pressure. It is evident that entropy is more temperature-sensitive than pressure-sensitive.

### 3.5 Magnetic properties

The Slater Pauling rule, $M_{total}=Z_{total}-18$ for calculation of total magnetic moment (MT) per formula unit for the given half-metals with C1b structures is applied. $Z_{total}$ stands for the total number of valence electrons. Therefore, based on **Table 4** data, we anticipate an $M_{total}$ of $1.10\mu_B$ and 1.22$\mu_B$ VPdGe and VPdSn alloys respectively. The gap that occurs in the minority spin states and the density of states (DOS) near the $E_F$ of the majority spin states are the sources of the negative moment. For VPdGe and VPdSn alloys the total magnetic moment ($M_{total}$) per formula unit is nearly an integral value of 1.10$\mu_B$ and 1.22$\mu_B$ respectively, further confirming the half-metallic character of these two compounds. We have observed that V metals from d block elements in both alloys contribute significantly in total magnetic moment whereas the participation from Pd, Ge and Sn atoms is negligibly small.

**Table 4:** Partial and total magnetic moment ($M_{total}$)/ formula unit for VPdZ (Z; Ge, Sn) Heusler alloys

| Magnetic Moment ($\mu_B$) | V | Pd | Z | $M_{total}(\mu_B)$ |
|---|---|---|---|---|
| VPdGe | 1.06225 | -0.04327 | -0.06748 | 1.10 |
| VPdSn | 1.1077 | -0.05276 | -0.06797 | 1.22 |

Examining the DOS and pDOS from **Fig. 4** of these alloys helps to explain the magnetic behavior. The asymmetry in the DOS profile between the minority and majority spins reflects the origin of magnetism. The pDOS provides an explanation for the V and Pd atoms opposite magnetic moment alignment in these combinations. In minority spin, the d states of Pd are below the Fermi level ($E_F$), while in majority spin, they appear above the $E_F$. Because of this, the majority spin of V atoms is primarily unoccupied, giving Pd a significant and opposing magnetic moment. The magnetic moment of Pd keeps on decreasing whereas that of V increases results the net magnetic moment.

### 3.6 Lattice dynamic study

It is crucial to investigate in order to forecast the dynamic behavior of the crystal lattice phonon band structure. The phonon dispersion curves for VPdGe and VPdSn alloys can be clearly comprehended with the help of schematic **Fig. 9(a), (b)**. Three atoms make up each of the material's unit cell in terms of crystal structure, which ultimately results in nine vibrational modes. Three of these vibrational modes known as acoustic modes are linearly scattered modes at the Γ point that originate from the in-phase atomic displacements of atoms at their basis. The remaining six modes are called optical modes; they originate from the out-of-phase atomic translations and show nearly flat dispersion near the gamma (Γ) point. In a similar vein, the degeneracy of two sets of optical modes close to the Γ point also clearly displays the crystal symmetry. By closely examining the phonon band structure of materials, we are able to predict their thermodynamic stability with high accuracy. The acoustic phonons, which exhibit great dispersion, are the primary donors to thermal current in the lattice, while the optical modes, which are characterized by low dispersion and group velocity, contribute very little to it. Some materials, such as thermoelectric and thermal insulators, are widely used in a variety of industrial and commercial applications. For best results, they need very low lattice thermal conductivity.

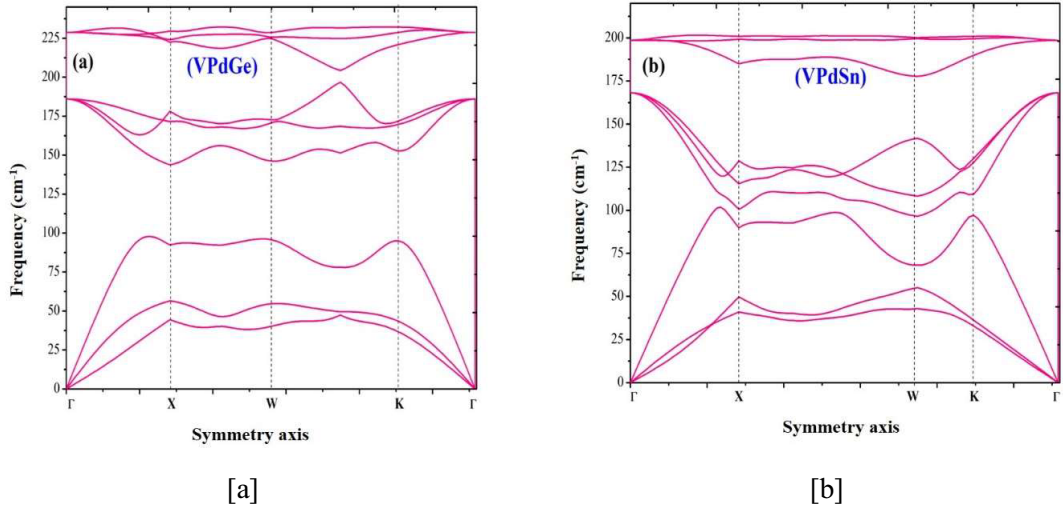

[a]                                                                 [b]

**Fig. 9:** Plots of phonon dynamic stability for [a] VPdGe and [b] VPdSn Heusler alloys

## 4. Conclusion

We have completed a thorough analysis of VPdGe and VPdSn half Heusler alloys and described their basic features in the work reported. The spin polarized arrangement and phase-α stability of these crystalline structures were confirmed by the density functional theory calculations. The half-metallic feature is indicated by the spin-polarized behavior that is revealed by the detailed profiles of the electronic band structures. According to the electronic band profiles both alloys are half-metallic, with indirect energy gaps in the spin down channel of 1.10 eV and 1.02 eV for VPdGe and VPdSn half Heusler alloys respectively. The magnetic moments estimated integral values are found to be exceptionally compatible with the Slater Pauling rule. Pugh's ratio (B/G) and other elasto-mechanical characteristics reveal that VPdGe is ductile whereas VPdSn is brittle in nature. The dynamic stability at various temperatures and pressure conditions are confirmed by their thermodynamic characteristics. Furthermore, the lattice dynamic studies validate the dynamic stability and strongly support the potential for their synthesis in future solid-state electronics and renewable energy devices.